# TRADING ON THE FLOOR AFTER SWEEPING THE BOOK

## Vassilis Polimenis*

*Informed traders need to trade fast in order to profit from their private information before it becomes public. Fast electronic markets provide such liquidity. Slow markets provide execution in an auction based trading floor. Hybrid markets combine both execution venues. In its main result, the paper shows that to compensate for their slow and risky executions, trading floors need to be at least twice as deep as the sweeping facility. Furthermore, when a stand-alone trading floor is enhanced with the addition of a sweeping facility, overall informed trading will decline because it is easier for informed traders to extract the full value of their private info.*

> The NYSE Hybrid Market will marry the best of electronic trading and the auction market in a way no other can match. — John A. Thain, NYSE CEO

Advances in communication and computing technologies are dramatically altering the trading landscape. The traditional trading paradigm, that of an "open-outcry" auction based trading floor, is giving its place to a more complicated and diverse order matching environment. As the introduction of new technologies brings promise for faster, less risky and more accurate trading executions, it is becoming increasingly difficult for traditional exchanges to compete for order flow. This difficulty is evident in the recent merger activity; during April 2005, all four major U.S. equity markets were seeking to merge.[1] Mergers between traditional and electronic exchanges will produce hybrid markets that combine manual and automatic execution elements.

Such transformations have, at the very best, been received with skepticism by the floor-trading communities of brokers and specialists — the intermediaries who oversee the matching of buy and sell orders — who are at the heart of the auction-based price-discovery process. In a decimalized world, complete automation that

---

1. On April 20, NYSE and Archipelago aanounced an agreement to merge and, furthermore, to become a public company. Two days later, NASDAQ announced an agreement to buy the INET ECN.

---

Vassilis Polimenis is an assistant professor in the A. Gary Anderson Graduate School of Management at the University of California. Contact information: University of California, A. Gary Anderson Graduate School of Management, 900 University Ave., Riverside, CA 92521-0203. E-mail: polimenis@ucr.edu

eliminates latency in order execution, and creates perfect order timing sequencing, will make it dramatically more difficult, or even impossible, for dealers and exchange professionals to trade profitably. As Peterffy and Battan noted in the SEC's Market Structure Hearings held in New York on November 2002, "[d]esignated liquidity providers, therefore, have had to rely on their inherent time and place advantage in the manual market place — specifically, that they can see orders before others can see them and can take their time (sometimes up to 90 seconds) to decide whether to interact with those orders or not — in order to reap a reward for the services they provide." (See also the recent Peterffy and Battan 2004 piece in *Financial Analysts Journal*.)

　　The key difference between traditional "slow" floor-based exchanges relative to their "fast" electronic competitors for liquidity is the firmness of quotes posted in slow and fast markets. A fast market quote is a true quote that an order can undoubtedly get filled against, while a slow market quote is more of an indication of a price. In fast-changing market conditions, a slow market may advantageously use its "slowness" in updating prices, thus posting artificially attractive quotes, and block away electronic exchanges that cannot do so.

　　To modernize the National Market System [NMS] to reflect advances in electronic stock markets and order routing, in March 2004 the SEC established the new Regulation NMS, which among other things differentiates between automatic (fast) and manual (slow) markets. According to the new ruling, fast markets are not allowed to execute an order at a price that is inferior to an electronic market's best price but may trade through better but non-immediately executable prices on slow markets. The rule takes away the biggest competitive advantage of traditional trading floors in the battle for liquidity. As a response, many exchanges are proposing enhancements to their electronic-trading platforms.

　　One of the highly contested issues for these newly emerging hybrid exchanges is whether to allow for so-called "sweeping of the book." Sweeping refers to the ability to electronically execute transactions not only at the best price, as for example NYSE's electronic platform Direct+ now does, but also at quotes above or below the best price. It is argued that the electronic component creates in-house competition for the trading floor, since electronic access to the full limit order book would make it much more challenging for the floor community to co-exist within the newly emerging hybrid markets.

　　In the face of these significant developments, the academic community has been slow in offering definite answers to the questions surrounding these issues. Are hybrid markets good for price discovery? Do trading floors enhance the efficiency of the market? What changes in trading should would we expect when a traditional exchange becomes hybrid by offering electronic access to its entire limit order book through sweeping? What are the parameters that will determine the inter-market competitiveness?

　　The main reason for the lack of definite academic answers is the inherent complexity of the trading process and the difficulty of modeling liquidity discovery. An important growing branch of literature endogenously explains liquidity constraints and their effect on trading. In the two canonical models of market microstructure described in Glosten and Milgrom (1985) and Kyle (1985), liquidity discounts are related to the size of the order because large orders are more informative. Recently, Back and Baruch (2004) provide some joint analysis of the Kyle and Glosten and Milgrom papers with a single informed trader and risk neutral market makers.

Despite their indisputable significance, the above equilibrium analyses have been criticized in that they do not lend themselves to real market calibrations, and their conclusions cannot be generalized to more realistic models of trading (Black 1995). For the discussion of slow versus fast markets, these models are silent because, in order to endogenously derive market depth, they posit efficient markets that reflect all publicly available information immediately.

In this paper we take some initial steps in the direction of modeling fast and slow markets and discussing their basic competitive characteristics. Slow markets, as their name implies, take some time to execute orders; fast markets are the limit of slow markets with infinite speed of adjustment. There is another important difference: In electronic markets limit orders are pre-posted and cannot be removed depending on the incoming order size; the electronic market does not have time to differentiate for small versus large traders. On the contrary, in trading floors, due to the slow speed of execution, market makers have the necessary time to adjust their quotes when a block is coming and may thus avoid to be price discriminated against (by not providing liquidity at the early stages of a block's execution).

The first contribution of the paper is to model market dynamics as a function of trading volume rather than the usual calendar time, t. When the market evolves stochastically as a function of volume, execution time of a size q order becomes a stochastic process with respect to q; for a particular execution scenario $w\hat{I}W$, a trade will be completed at time $t = t(q, w)$. Price dynamics are then subordinated to execution time dynamics in a way that will formally be defined in the text. The value of trading volume in explaining returns is being discussed in Clark (1973), and Ané and Geman (2000).

The paper argues that hybrid markets do not generate more informed trading than stand-alone electronic markets. That is, under the assumptions of the model here, a slow market only competes for the same liquidity with the fast market. It is shown that, when competing for liquidity with an electronic (Kyle-type) market, which is $l^{-1}_F$ shares per dollar deep, a trading floor will be able to divert trades away from its electronic competitor only if it is at least twice as deep.

Actually, at its concluding section the paper shows that when a stand-alone floor is transformed to a hybrid market, with the addition of a sweeping facility, we could expect the overall amount of informed trading to decline. The fundamental reason for such a decline is that information mainly flows from the fast to the slow market; thus, it is the lower depth of the fast market that determines overall trade.

The first task of the paper is to present a formal model of the liquidity and price discovery mechanisms on both the trading floor as well as the hybrid markets that combine a floor with an electronic sweeping facility. The model presented here, using only mild assumptions, builds upon the basic notions of liquidity — order size that can be absorbed in a given time — and depth — units of liquidity for a given cost — as market participants truly understand them. For a short period, before her proprietary information becomes public, an informed trader is a liquidity monopsonist. In return for releasing information in the market, through her trades, the informed trader is compensated with liquidity supplied by uninformed traders.

Unlike the instantaneous and riskless execution offered by electronic markets, auction-based trading floors are characterized by "slow" executions. In a related study, Polimenis (2005) investigates the optimal trading of a trader who may only trade in a slow market. The central idea in that paper, which we borrow here, is that in a slow market price adjustment is subordinated to liquidity discovery; prices are updated while we search for liquidity. Thus, a slow order execution is undesirable for an informed trader, not due to lost interest in the money "tied" to the order but, rather, due to informational dimension of time. The motivation behind Polimenis (2005) is to capture the fact that a slow execution is risky because it exposes the block order to the market for prolonged periods of time and thus carries informational costs. As is discussed there, execution time provides the medium for incorporating new information in prices.

As a side note, it may be argued that such an approach is not only motivated by purely practical reasons but in a fundamental way relates to the complexity of computing the efficient price by processing in real time trading patterns and other information available to market makers. By prolonging executions, illiquid markets allow more time to the market to absorb and reflect private information, thus limiting the informed agent's profits.

The model of the slow market here is related to Polimenis (2005), but here the trader's options are extended to allow for the choice to participate, up to an endogenously chosen degree, in a fast market. Another difference between the two models of slow markets is, in Polimenis (2005), noise trades arrive continually as a Brownian motion, while here, noise orders arrive at discrete but totally random sizes and times.

Notably, the applicability of the results is limited in that the paper does not endogenize the entire liquidity markets but rather only addresses in an endogenous fashion *inter-market* competition. The paper thus builds upon the groundbreaking work by Kyle and others, who provide a purely endogenous lambda, by enhancing the available trading venues.

In Sections I and II, the fast and slow markets are introduced. In Sections III and IV, we model liquidity discovery and optimal trading on the stand-alone floor. In Section V hybrid markets are discussed. Finally, in Section VI, the nature of the market for liquidity is discussed.

## I. FAST MARKETS AND KYLE'S LAMDA

Modeling the operation of fast markets is relatively straightforward. Fast markets (typically electronic order books) are characterized by certain and immediate executions; that is, when sweeping a transparent book, the agent knows all dimensions of her order execution quality. The order will be immediately executed at a price away enough (from the current price) to activate the required amount of standing limit orders.

The most important parameter that determines execution quality in a fast market is the density of standing limit orders. Such markets are characterized by their breadth, $\lambda_F$ (F stands for Fast market), measured in *dollars* of expected impact *per transacted share*. That is, $\lambda_F$ is the sensitivity of the fast market to trading. Kyle (1985) endogenously calculates the $\lambda$ of the market as a signal-to-noise ratio. The trader who trades q shares in a fast market with $\lambda_F$ will experience a deterministic adverse impact

$$\lambda_F q \tag{5}$$

The inverse quantity $\lambda_F^{-1}$ is the familiar market depth, measured in *shares per dollar,* with the natural interpretation of capturing the density of shares, placed by limit orders, per dollar in the order book. Clearly, breadth $\lambda_F$ is a negative attribute since it measures the cost per unit of liquidity. Depth measures the liquidity that one dollar buys in the market.

The value of private information is measured by the implied *misvaluation* from their fundamental value at which securities currently trade. According to her private information, the trader observes that, at time zero, the stock trades at a price that deviates from its intrinsic value by $\Delta P$ dollars. If the trader were a price taker, the existence of a non-zero $\Delta P$ would be an arbitrage opportunity, and thus an unacceptable condition for an infinitely deep market.

In reality, even though the condition $\Delta P>0$ will certainly lead to trading, it does not amount to an arbitrage opportunity. To remain profitable, the trader has to balance the amount of trading with the impact it generates. In this environment, the profitability of her strategy is determined by the value of the private information, $\Delta P$, which captures the gain due to the current misvaluation of the security, and the price impact of the released information (liquidity cost), I.

## A. Sweeping the Book

When sweeping q shares from the book, the agent releases information that immediately impacts the price by $\lambda_F q$. It should be emphasized here that limit orders on the book are already standing and the instantaneous nature of order execution in fast markets implies that *sweeping the book is an inherently anonymous operation.*[2] That is, the fast market here differs from the typical one-shot model (e.g., Kyle) where a unique price that clears the market applies for the entire order. When the insider chooses to sweep the book, her orders are being executed at progressively deteriorating price points. Thus, the total cost of buying the required liquidity from the book equals $\int_0^q \lambda_F y\, dy$, and her profit is

$$q\Delta P - \int_0^q \lambda_F y\, dy \tag{2}$$

---

2. This will be contrasted later with trading floors that do not provide anonymous trading.

The trader will trade $q_F$ shares, where

$$\Delta P = \lambda_F q_F \tag{3}$$

After $q_F$ shares the price reflects the full private information and trading stops. Thus, a fast market with instantaneous executions will immediately return to its efficient state even in the presence of a single informed trader.

## II. TRADING IN SLOW MARKETS

Slow markets, as their name implies, are characterized by slow and thus uncertain executions; that is, when sending her order to a trading floor, the agent does not know when her order will be executed and at what price. Observe that if, as in Kyle, the trader is risk neutral, risk and thus execution speed don't matter.

Since trading floors are risky, in order to attract traders, they have to provide, in a sense that will become clear, "cheaper" liquidity than books. In liquid trading floors, the large number of uninformed traders allows an informed trader to release less information per share traded. In this sense, the model of a slow market here resembles Kyle (1985), but with an important difference: In Kyle market makers decide an *efficient price* by completely "distilling" all public information. Here, market makers still correct prices in the right direction (i.e., towards efficiency), but not in a completely efficient way. Instead, they *mechanically tend* to correct prices upwards for a buy, and downwards for a sell. Even though this implies that prices are to some degree inefficient, it resembles more with realistic markets, where market makers may not want to impose a strong structure on the missing inside information (i.e., on what they don't know).[3]

Since liquid markets are characterized by the large number of shares that can be traded before private information becomes public, it is natural to define floor liquidity, *a*, as the expected rate at which marketable orders arrive; and thus measure it in <u>shares per second</u>.

In a way analogous to the lambda of a book, $\lambda_F$, a floor is also characterized by its sensitivity to new trades, $\lambda_S$ (S for Slow market). The fundamental difference between the sensitivities of a fast and slow market, is that the latter refers to a risky execution and thus needs to be corrected for risk.

Unlike a fast Kyle-type market that adjusts prices immediately to reflect new information, the $\lambda_S$ of a slow market is only an expected rate of adjustment. In a slow market, prices are expected to adjust at a rate

---

3. In Kyle, the endogenous $\lambda$ depends on knowledge of the variance of the inside signal, and in the more recent version in Back and Baruch (2004), it depends in restricting the two possible inside values (zero or one).

$$\dot{I} = \ddot{e}_S \, a \qquad (4)$$

*dollars per second.*

A proper model of slow markets should include the fast market as a special case for which $a = \infty$. When liquidity is infinite, there is effectively no search for liquidity, and prices adjust immediately, $\mu = \infty$. Furthermore, if two markets are characterized by the same $\lambda$, the one with the higher $a$ is superior and will be strictly preferred by the agent. This is because, as we will see later, a higher liquidity, $a$, implies faster execution and thus smaller execution risk.

The agent who observed that the illiquid security was last traded at a price away from its actual value by $\Delta P$ dollars knows that if she chooses to trade she will not enjoy the entire $\Delta P$, since information will affect the market in an adverse way. For example, a large buy order (for an undervalued security) will tend to get executed at a higher transaction price. Thus, the trader will pocket a per share profit equal to $\Delta P - I(q)$, where $I(q)$ is the total dollar impact of information released during the time, $\tau(q)$, it takes for the q-order to get executed.

## III. LIQUIDITY DISCOVERY ON THE FLOOR SATISFIES

In order to model a slow market, we have to first model the process of liquidity discovery on the floor. Generally speaking, execution on the floor is stochastic, and determined as a particular realization $\omega$ from a universe $\Omega$ of possible execution paths. Under a particular execution scenario $\omega \in \Omega$, a trade will be completed at time $\tau(q,\omega)$ when enough traders willing to take the opposite side have submitted their marketable orders. If we make the natural assumption that, for every execution $\omega$, larger orders will take more time to get executed, we may think of execution time, $\tau$, when plotted against size, q, as an increasing process taking values in $(0, \infty)$. If we further assume that the extra time it takes to execute another q' shares does not depend on how many shares q the trader has executed so far — formally $\tau(q+q')-\tau(q)$ is distributed as $\tau(q')$ — we are left with $\tau(q)$ being an increasing Lévy process in q.[4] In the language of Lévy processes, $\tau(q)$, when viewed as a function of size q, is a subordinator.[5]

An important feature of Lévy subordinators is that we may work with their cumulant kernel $K_\tau()$, defined for any s through

$$Ee^{s\tau} = \exp(qK_\tau(s)) \qquad (5)$$

The increasing process t(q) is clearly of bounded variation with no negative jumps or diffusive component. The rate of arrivals of *execution-time jumps* with size between z and z+dz, the Lévy measure L(dz), in this case satisfies

---

4. We are used to processes with respect to time; here time (and price impact) is a process with respect to q.
5. For more on subordinator Lévy processes and their generating functions see ch.III in Bertoin (1996).

$$\int_0^{+\infty} \min(1,z) L(dz) < \infty \tag{6}$$

In the most general case, a subordinator includes a *locally deterministic* continuous and a pure jump part. In the case of the τ(q) process, given that order arrivals from noise traders in reality are totally random in size and unpredictable, it makes little sense to include a deterministic continuous component. Thus, we will assume here that τ(q) includes only positive jumps[6]; in this case the Lévy-Kintchine representation of the kernel is given by

$$K_\tau(s) = \int_0^{+\infty} (e^{sz} - 1) L(dz) \tag{7}$$

Then, the extra time it takes for an extra dq shares to get executed follows

$$d\tau = \int_0^{+\infty} z N(dq, dz) \tag{8}$$

where, the law of N is that of a Poisson random measure on $\Re_+ \times \Re_+$ with Lévy measure L(dz) that models the pure jump liquidity arrival delay from the totally random market orders submitted by noise traders. That is, N(dq,dz) is one when it takes between z and z+dz seconds for an extra dq shares to get executed. This means that an incremental infinitesimal quantity may take a large amount of time to get executed.

Since liquidity is formally measured as the expected rate, *a*, at which marketable orders arrive, and is measured in transacted shares per second, the trader expects that her order will take Eτ = q/a seconds to get executed on the floor. On the other hand, from the Lévy property we have that

$$E\tau = q K_\tau'(0) \tag{9}$$

which implies that liquidity has to satisfy

$$a^{-1} = K_\tau'(0) = \int_0^{+\infty} z L(dz) \tag{10}$$

**Example.**

In this example, provided to make the discussion specific, we specify the exact dynamics of the execution times τ as a process of order size q. Example equations retain the original label with an e-suffix. As we discussed already, τ(q) is a subordinator with respect to q. The *Gamma process* provides an important family of subordinators

$$\tau = \gamma(a^{-1}, q) \tag{11}$$

---

6. The inclusion of a deterministic part is analytically trivial.

In this case the distribution of the execution time τ for an order of size q is

$$P(\tau(q) \in d\tau) = ae^{-a\tau}(a\tau)^{q-1}/\Gamma(q)d\tau \quad \text{for } \tau > 0 \tag{12}$$

The Gamma process evolves purely through totally unpredictable jumps, and in this case (7) specializes to

$$K_\tau(s) = -\log(1 - s/a) \tag{7e}$$

The idea of subordinating price impact to the liquidity discovery is introduced for the first time in Polimenis (2005), and our slow market here strongly parallels the trading floor in that paper. One difference between the two models, besides the fact that here the slow market will compete with a fast market for liquidity, is that in Polimenis (2005) liquidity discovery is purely continuous, and the execution point in (11) is defined as the stopping time when a Brownian motion meets the level q. As a consequence in Polimenis (2005), the execution time follows an Inverse Gaussian law.

Given (8), the uncertainty in execution delays is given by

$$\text{Var}(\tau(q)) = q \int_0^{+\infty} z^2 L(dz) = qK_\tau''(0) \tag{13}$$

**Example, continued.**

Given (11) and (13), the variance of execution delays for the Gamma execution follows

$$\text{Var}(\tau) = q/a^2 \tag{13e}$$

Uncertainty grows with size and declines with the liquidity of the market.

Having modeled *liquidity discovery,* the next issue is that of the evolution of prices during execution, that is, *price discovery.* The uncertain execution time and the uncertain price evolution during execution introduce execution risk in slow markets. If the price discovery process was deterministic, as in a fast market, the informed trader would know that when her order execution would be completed, her per share profit would be exactly equal to $\Delta P - \lambda_s q$.[7]

In reality, $\mu$ in (4) provides only an expected rate at which the price drifts "against" the trader. To capture the stochastic nature of real markets, impact is modeled as a Brownian motion with drift $\mu > 0$.

$$I(q) = \mu \tau(q) + \sigma W(\tau(q)) \qquad (14)$$

where $\tau(q)$ is the stopping time defined above. Furthermore, the impact Brownian motion W is independent of noise trade arrivals. Even though price impact evolves as a Brownian motion with respect to execution time, when impact is studied as a function of size, which is the true control variable for the informed agent, it does not obey a Gaussian law anymore.

In real markets price discovery continues while the floor searches for liquidity, and this is captured here by the mathematical *subordination* of the price discovery to the liquidity discovery process in (14). Taking iterated expectations, by conditioning on the delay $\tau$ and using the moment generating function for $\tau$, we recover

$$E^q e^{sI(q)} = E^q E^\tau e^{sI(\tau(q))} = \exp(qK_\tau(\mu s + .5\sigma^2 s^2)) = \exp(qK(s)) \qquad (15)$$

so that the cumulant generator of the impact process is related to the cumulant kernal of the delay through

$$K(s) = K_\tau(\mu s + .5\sigma^2 s^2) \qquad (16)$$

**Example, continued.**

In the case of the delay process in (11), given (7e), the impact follows a *Variance Gamma* (VG) process.[8] The cumulant function K(s) in this case equals

$$K(s) = \log(a/(a - \mu s - .5\sigma^2 s^2)) \qquad (16e)$$

We already know that the agent expects that the impact of her order will be

$$EI_S = \lambda_S q = K'(0) q \qquad (17)$$

From this we find the first derivative of the impact cumulant function at zero

$$K'(0) = \lambda_S \qquad (18)$$

---

7. And this would trivialize the issue since the trader would just choose to trade the entire quantity on the low $\lambda$ market.
8. The VG process (with respect to time) is introduced in Madan and Milne (1991), Madan and Seneta (1990), and Madan, Carr, and Chang (1998).

The expected impact may also be recovered from (16) by noting that

$$K'(s) = (\mu+\sigma^2 s)\, K_\tau'(\mu s+.5\sigma^2 s^2)$$

which, using (10), leads to

$$\lambda_S = K'(0) = \mu\, K_\tau'(0) = \mu\, a^{-1}$$

which is a proof of the originally desired relation (4). Execution price risk equals

$$Var(I) = K''(0)\, q \qquad (19)$$

Given (16), we have

$$K''(0) = \mu^2\, K_\tau''(0) + \sigma^2\, K_\tau'(0) \qquad (20)$$

or more intuitively

$$Var(I) = Var(\tau)\,\mu^2 + E(\tau)\,\sigma^2 \qquad (21)$$

Cumulant functions are always convex, $K''(0)>0$, so that a large order will increase the price impact and its variance. Liquidity lowers risk because it affects the execution delay.

**Example, continued.**

When execution delays follow a Gamma process, $Var(\tau)$ follows (13e), and (21) becomes

$$Var(I) = (\lambda_S^2 + \sigma^2/a)q \qquad (21e)$$

Notice that, keeping $\lambda_S$ fixed, *liquidity does not affect expected price impact.* Nevertheless, a high-*a* security will be preferred because, by accelerating execution, *a* lowers price risk.

## IV. TRADING ON THE FLOOR

The trader exhibits a constant absolute risk aversion $\eta$ and knows the liquidity parameters of the market; i.e. knows the liquidity offered by the floor, *a*, and the $\lambda$ of the floor, so that she can use (4) to calculate the impact drift $\mu$ for her orders. She also knows the volatility parameter for the security, $\sigma$.

By offering slow executions, a trading floor has another important disadvantage for the insider; knowing it will impact the price, market makers who observe a large incoming order will refuse to provide the crucial early liquidity and be price discriminated against. Unlike fast markets, where the instantaneous execution allows the trader to "hit" limit orders at their pre-committed levels, in slow markets, market makers may change their quotes after having seen the order. As was discussed in the introduction, this is really the most crucial point in the entire debate about slow and fast markets. In other words, here, the slow market is a one-shot model in the sense that the entire order gets executed at a single price.

When the order is submitted to the trading floor, the block will be transacted at a single price that fully reflects the entire impact of the information released during the order execution. The utility — a trader with initial wealth W gets — from issuing a block order of size q equals

$$U_S = E^q -\exp(-\eta (W + q(\Delta P - I_S (q)))) \qquad (22)$$

The q-superscript in the expectation operator explicitly shows that the trader is not a price taker; that is, the order size will determine not only the trader's position but the price impact as well. That is, the expectation is taken *conditionally on the chosen order size*. Observe that, for the analysis of the optimal order, the direction of the mispricing does not matter, since it will only determine the direction of the trade (buy for underpriced and sell for overpriced securities). Without loss of generality, we may assume that $\Delta P$ and q are positive.

We find that the utility gain[9] from block trading in a slow market equals

$$G^S = \log (U_0/U_S) = \eta q \Delta P - q P_q^S \qquad (23)$$

with the price of floor liquidity being given by

$$P_q^S = K(\eta q) = K_\tau(\mu \eta q + .5\sigma^2 \eta^2 q^2) \qquad (24)$$

**Example, continued.**

Given the Gamma dynamics, the price of liquidity in (24) specializes to

$$P_q^S = \log(a/(a - \mu \eta q - .5\sigma^2 \eta^2 q^2)) \qquad (24e)$$

Observe that, with this interpretation of $P_q^S$ as the price of liquidity, and since $K(0) = 0$, we get the proper result that for small traders (price takers) the price of liquidity $P_0$ is zero; a small investor is a price taker since she does not have to wait for liquidity.

When trading on the floor, as the sole owner of her "aging" proprietary information, the trader is faced with a dilemma: trade a small quantity (quick) thus capturing the maximum per share benefit, or trade a large quantity (slow) at smaller per share gain. The informed trader behaves as a time-limited *liquidity monopsonist* confronted with what essentially amounts to an increasing supply curve for liquidity[10]

---

9. Exponential utility is negative, and thus the agent is better off by actually lowering his utility in absolute terms.
10. Clearly, since $K'(0) = \lambda_S > 0$, and given the convexity of cumulant functions $K(s)$, $dP/dq > 0$.

$$dP/dq = K_\tau'(\mu\eta q + .5\sigma^2\eta^2 q^2)(\mu\eta + \sigma^2\eta^2 q) > 0 \qquad (25)$$

**Example, continued.**

Given Gamma dynamics, the price derivative specializes to[11]

$$dP/dq = (\mu\eta + \sigma^2\eta^2 q)/(a - \mu\eta q - .5\sigma^2\eta^2 q^2) > 0 \qquad (25e)$$

**A. The Optimal Order on the Standalone Floor**

In slow markets, the informed agent's problem is to decide for the optimal block order size. Given (23), a large trader will choose a size q that satisfies

$$q_S = \mathrm{argmax}_q \; \eta q \Delta P - q P_q^S \qquad (26)$$

As a liquidity monopsony, the informed trader trades to the point that equates her constant marginal revenue

$$MR = \eta \Delta P = AR \qquad (27)$$

to her increasing marginal liquidity cost

$$MC = P_q^S + q dP/dq > AC = P_q^S \qquad (28)$$

The optimal trade point $q_s$ is shown in Figure 1.

**V. TRADING ON THE FLOOR AFTER SWEEPING THE BOOK**

In a hybrid market, where a trader has access to a trading floor as well as the entire limit order book, she has one more control variable. Namely, she can choose how many shares to sweep from the book, $q_F$, and send her remaining order, $q_S$, to the floor for execution. Notice that we do not assume here any ex ante preference for one market over the other; *the order split choice and submission to both markets happens simultaneously.* Nevertheless, since sweeping the book is deterministic and instantaneous, while trading on the floor is slow, the book trade will materialize before the floor trade has even started. This important observation is central in the entire debate around slow versus fast markets; due to their ability to instantaneously reflect it, information flows easier from fast to slow markets rather than the other way around. Some who, probably unfairly, argue that trading floors profit on the expense of electronic markets imply this kind of consideration.

As we saw previously, when a block order is submitted to a floor, the block will be transacted at a single price that fully reflects the entire impact of the information released during the order execution. If $q_H = q_{HF} + q_{HS}$ is the entire trade size in a hybrid market, then the ratio $q_{HF}/q_H$, denotes the fraction of her order the trader chooses to sweep the book for and thus captures the *immediacy* of the order. Equivalently, the immediacy factor $q_{HF}/q_H$ determines how deeply (i.e., how far from the current market) she sweeps the book.

---

11. Clearly, the trade quantity q will always be such that $a - \mu\eta q - .5 \acute{o}^2\eta^2 q^2 > 0$; otherwise, the trader will incur an unbounded liquidity price (24e).

**Figure 1.**

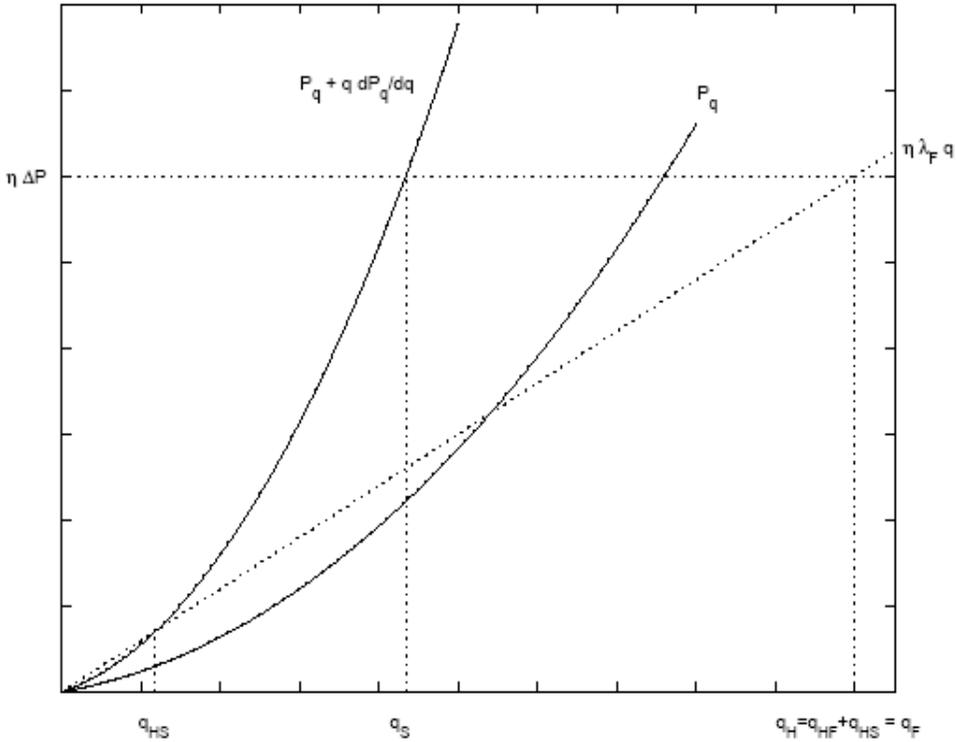

The total impact of a hybrid trade is decomposed into two different sources: (1) a deterministic component because of sweeping and (2) a stochastic due to private information leaking and becoming public during execution on the floor.

Thus, the total utility gain from trading in a fast electronic market, and a slow and transparent trading floor equals

$$G^H = \eta q_H \Delta P - \eta \int_0^{q^F} \lambda_F y \, dy - \eta q_S I_F - q_S P_{qS} \qquad (29)$$

Essentially, the trader buys the liquidity to trade in two different places. Sweeping the book happens instantaneously, before the floor has had any time to incorporate information, and thus the liquidity cost of the fast component in a hybrid market remains unaffected as if the fast market where to operate alone (not as part of a hybrid market), $P_q^F = \eta \lambda_F q$.

On the other hand, since sweeping the book releases information immediately, this information will also be immediately reflected in trading ensuing on the floor. Essentially, market makers on the floor observe the electronic component and adjust prices accordingly. Thus, the liquidity price in the slow market (trading floor) will also depend on the amount traded on the fast market (limit order book). This explains the term $\eta q_S I_F = \eta q_S \lambda_F q_F$ in (29) above.

Taking the derivative of (29) with respect to the amount of total trading, we find that the amounts traded in the fast and slow markets have to combine as follows

$$\Delta P = \lambda_F q_H = \lambda_F (q_{HF} + q_{HS}) \tag{30}$$

But from (3), we see that a stand-alone fast market would provide the same result,

**Lemma 1.** Hybrid markets do not generate more informed trading than stand-alone fast markets.

In other words, the total trading in the hybrid market is determined from the $\lambda$ of the fast component. When it co-exists with a fast market, a slow market "steals" liquidity without offering any informational benefits.

To some degree, lemma 1 is a negative statement for the existence of the so-called hybrid markets that combine electronic and auction based execution facilities. But we have to be careful in that lemma 1 is only valid under the restricted trading policies considered here, that is, one-shot policies. That is here, we have not considered the fully dynamic problem where the trader trades small quantities, and as the impact of her previous sub-orders gets realized, she decides how to trade next.

Finally, lemma 1 predicts something many market participants have asserted all along: The two types of markets are <u>inherently competitive</u> since they will have to share the same liquidity. Even though the sensitivity of the slow market will not affect the overall informed trading, it will play a central role in determining the degree to which the trading floor will be able to compete in "stealing" liquidity away from the book.

Having solved for the optimal $q_H$, rewrite (29) as a function only of the trading on the hybrid floor, $q_{HS}$

$$G^H = \eta q_H \Delta P - \eta \int_0^{q_H - q_{HS}} \lambda_F y\, dy - \eta q_{HS} \lambda_F (q_H - q_{HS}) - q_{HS} P_{q_{HS}} \tag{31}$$

and take the derivative of (31) with respect to $q_{HS}$

$$\frac{\partial G^H}{\partial q_{HS}} = \eta \lambda_F q_{HS} - P_{q_{HS}} - q_{HS} \frac{dP_{q_{HS}}}{dq} \tag{32}$$

The optimal trade point $q_{HS}$ is clearly shown in Figure 1 as the point where the line with slope $\eta\lambda_F$ meets the MC = $P_q^S$ + qdP/dq curve.

Since this partial at zero is zero, the trader will only trade at the floor when the second derivative at zero is positive. Taking the second derivative of $G^H$, with respect to floor trading, at zero, and observing from (25) that

$$[dP/dq]_0 = K_\tau'(0)\mu\eta = \eta\lambda_S \quad (33)$$

we find that

$$\lambda_S < \lambda_F/2 \quad (34)$$

**Lemma 2.** In a hybrid market, trading will be diverted from the fast to the transparent slow market only if the lambda of the floor satisfies

Lemma 2 formalizes and quantifies the previous informal reasoning that slow markets, being risky, will have to somehow provide cheaper trading to attract liquidity. Indeed, criterion (34) points to cheaper trading, since the trader expects less than half the impact per share. Lemma 2 shows that the proper dimension at which competing markets are measured is depth, $ë^{-1}$, rather than liquidity alone.

## VI. THE NATURE OF THE MARKET FOR LIQUIDITY

It is easy to see that as a liquidity monopsonist, the informed trader buys less liquidity thus revealing less information in the markets. But how much informed trading, $q_{max}$, can we expect at the best? What is the maximum profitable amount of trading from a social point of view? The value of the private information has been completely collected when there are no more utility gains from re-distributing the private information.

Specifically, this is the point at which a new trader, endowed with the insider's information, cannot profitably trade.

Let's start with a market where informed traders have already submitted trades for q shares. Since a new trade will have to be executed in a floor that already works on executing market orders for q shares, the post-trade utility for a new trader j who has been endowed with the information equals

$$U(q_j) = U_o \, E^{qj} \exp(-\eta q_j \, (\Delta P - I(q+q_j))) \quad (35)$$

The critical observation is that, as a subordinated Brownian motion, price discovery in (14) is a Lévy process with respect to trading q, not time; that is, $Ee^{sI(q)} = e^{qK(s)}$. A Lévy process is characterized by increments that are independent and identically distributed. Thus, given the trade q, the utility gain equals

$$G^j = \log \frac{U_o}{U(q_j)} = \eta q_j \Delta P - (q + q_j) P_{qj} \quad (36)$$

Equation (36) shows that our model captures the following fundamental, and, at first, counter-intuitive characteristic of the market for liquidity. In commodity markets, the price of the commodity is determined by the total demand, and the total cost for the $j_{th}$ agent equals the product of her demand times the market-wide clearing price, $q_j P_{q+qj}$. In the market for liquidity, *each agent pays an individual* *price* totally determined only by their trade $P_{qj}$, but their total cost depends on the entire liquidity demand, $q+q_j$. This happens because, liquidity markets differentiate agents by the amount they trade. Unlike "small" traders, who pay little as $P_0=0$, agents who trade large quantities are informed and they pay a large price.[12]

From (36) the marginal gain of the new trader is

$$?G_j/?q_j = \eta\Delta P - P_{qj} - (q+q_j)\, dP_{qj}/dq_j$$

Since, at $q_{max}$, no trader can benefit if given the information, the maximum amount of trading is such that the initial marginal gain *for a new trader*[13] has to be zero

$$\eta\Delta P - q_{max}\,[dP/dq]_0 = 0 \qquad (37)$$

where, we used the fact that $P_0 = 0$. From (33) we have

$$[dP/dq]_0 = \eta\lambda_S$$

and we recover

$$q_{max} = \Delta P/\lambda_S \qquad (38)$$

When contrasted with (3) and (30), we observe that eventually a stand-alone trading floor will reach the competitive point (38), but that will only happen when the private information leaks to an increasing number of traders.

Equation (30) points to another result. When a stand-alone floor (type S) is enhanced with a fast execution component to become a hybrid market (type H), overall informed trading is expected to decline. This decline happens because, from lemma 2, the ë of the fast market is quite higher. Since the overall trading in the hybrid market is determined by the $ë_F$, and not the $ë_S$ anymore, total trading declines. Essentially, it will be easier for many informed traders to extract the maximum value of their private information and they will trade less. Of course, if there is only a single informed trader, the introduction of the book sweeping facility will lower trading on the floor but increase overall informed trading as is clearly shown in Figure 1.

## CONCLUDING REMARKS

The paper introduces a formal model of the liquidity and price discovery mechanisms in a hybrid market, which combines a trading floor with an electronic sweeping facility. We show that, despite some common beliefs, hybrid markets do not generate more informed trading than stand-alone electronic markets. Under the assumptions of the model here, a slow market may only divert away liquidity from a fast market when it is at least twice as deep. Hybrid markets improve stand-alone trading floors, when the informed trader is the sole information owner. On the contrary, when there are multiple informed traders, transforming a trading floor to a hybrid market may lead to a decline of the overall amount of informed trading.

---

12. Clearly, anonymous markets, which we don't study here, to the benefit of informed traders cannot differentiate them by their size. Such markets are discussied in Polimenis (2005).
13. Since traders here are risk averse, the maximum trade will happen only by a new trader who is not already exposed to risk.